\documentclass[preprint,aps]{revtex4-1}
\usepackage{graphicx}
\usepackage{dcolumn}
\usepackage{bm}
\usepackage{amsmath}
\usepackage{amssymb}
\usepackage{latexsym}
\usepackage{epsfig}
\usepackage{amsbsy}
\usepackage{array}
\usepackage{setspace}

\begin{document}

\title{The higher-order black-body radiation shift of atomic energy-levels}

\author{Wanping Zhou}
\affiliation{School of Physics and Technology, Wuhan University, Wuhan 430000 China}

\affiliation{Engineering and Technology College, Hubei University of Technology, Wuhan 430000 China}
\author{Xuesong Mei}
\affiliation{School of Physics and Technology, Wuhan University, Wuhan 430000 China}
\author{Haoxue Qiao \footnote{Haoxue Qiao; electronic mail: qhx@whu.edu.cn}}
\affiliation{School of Physics and Technology, Wuhan University, Wuhan 430000 China}

\begin{abstract}
 
The one-loop correction and two-loop contribution to black-body radiation (BBR) shift are restudied. The S-matrix approach and nonrelativistic quantum electrodynamics (NRQED) are adopted in finite temperature case. The relativistic correction to one-loop BBR-shift has a $(Z\alpha)^{2}\alpha T^2/m$-order contribution. In the two-loop case, the pure thermal (real) photon part is too tiny to be detected; while the corrections induced by the thermal and virtual mixing diagram are at $(Z\alpha)^{2}\alpha^2 T^2/m$ order. We calculate the relativistic correction to one-loop BBR-shift in the ground state of hydrogen and ionized helium, which is larger than the leading term. As the leading term is proportional to $T^4/Z^4$. We estimate these higher-order corrections may be larger than the leading term, when the system is a highly ionized (large $Z$) or a cold (small $T$) one.

\end{abstract}


\maketitle

\section{Introduction} 
The interaction between blackbody radiation (BBR) and atomic system has long attracted physicists' interests. With rapid improvements in atomic physics, the importance of BBR is more and more obvious, in both fundamental and applied science. In determination of frequency and time, the roles of uncertainty induced by BBR were stressed in the most accurate clock experiments \cite{Chou00,Huntemann00,Madej00,Itano00}. By monitoring the BBR environment, the accuracy could also be significantly improved in optical lattice clock \cite{Bloom00}. The magnitude of uncertainties brought by the BBR-shift is $1Hz$ in the optical lattice clocks \cite{Porsev00}. It was the obstacle in determining the frequency standard at the accuracy of $10^{-18}$. This has stimulated a lot of works on calculating BBR-shift in the clock system \cite{Porsev00,Safronova00,Angstmann00}.

 On the theoretical work, Gallagher and Cooke first indicated that one should take BBR into account in Rydberg atoms even at room temperature \cite{Cooke00}. The first accurate calculation of dynamic stark shifts and depopulation rates of Rydberg energy levels induced by BBR was given by Farley \cite{Farley00}, whose work is one of the most important founding efforts on theoretical scheme allowing people to calculate BBR effects in an atomic system. The early works mainly focused on the Rydberg atoms since the atoms in such states have very large radii so that it strengthens the interaction between atoms and BBR. However, as the advance of high-precision spectroscopy and frequency metrology of atomic physics, the BBR effect on low-lying energy level transition is also increasingly showing its unique charm \cite{Jentschura00}. This will definitely lead a progress for understanding fundamental physical constants, such as Rydberg constant $R_\infty$ and fine-structure constant $\alpha$. Even more delighting, the Boltzmann constant, which is irrelevant with frequency transition, could be determined by a measured molar polarizability and calculation to the thermal effect brought by BBR \cite{Jentschura02}. In the last decades, tremendous theoretical schemes have been developed and used to help us to reveal and check the various subtle effects in atoms, such as QED-nuclear recoil (QED:quantum electrodynamics), QED-nuclear size, even if the weak interaction \cite{Eides00}. In the aspects of fundamental interests, a first-principle approach is naturally expected to be developed to allow us considering the effect like QED-BBR and so on. Based on the finite-temperature quantum field (FTQFT) approach \cite{Donoghue00}, Solovyev and the colleagues made an effort to derive a QED scheme to BBR effect on atoms \cite{Solovyev00}. In their work, a unique mixing effect which is from the field quantitation was reported (an additional correction to the level width). Escobedo and Soto obtained the leading term of BBR-shift by using the potential nonrelativistic quantum electrodynamics \cite{Escobedo00}. The higher-order BBR-shift is seldom reported. Jenschura and his colleagues \cite{Jentschura01} found some terms of BBR-shift in the P state of hydrogen-like atoms are proportional to $(ZT)^2$. However, the method they applied is adding the fine-structure and Lamb-shift to the formula of the BBR-shift by hand. In next sections, we would show that this $(ZT)^2$ term is related with higher-order one-loop BBR-shift and the leading term of two-loop BBR-shift.

In this work, we derive these higher-order corrections of BBR-shift by using the first principle method, which is the combination of the bound-state S-matrix \cite{Lindgren00,Sucher00}, finite-temperature quantum field theory \cite{Donoghue00} and non-relativistic quantum electrodynamics (NRQED) \cite{Jentschura01,Pachucki00}. The higher-order corrections here include the relativistic and quadratic corrections to one-loop BBR-shift, the thermal two-loop BBR-shift and the mixing two-loop BBR-shift, where the first three are induced by just thermal (real) photon, the mixing contribution is induced by virtual and thermal photon. The relativistic corrections to one-loop BBR-shift are the $(Z\alpha)^2\alpha T^2/m$ terms, and the mixing contributions are $(Z\alpha)^2\alpha^2T^2/m$ terms. They are the energy shift we mentioned before. We calculate the relativistic correction to one-loop BBR-shift in the ground state of hydrogen and ionized helium, which is larger than the leading term. According to our evaluation, in cryogenic environment or high-Z system (Z is the nuclear charge number), these next-leading-order corrections, which were obtained by first-principle, may contribute more importantly than the leading-order BBR correction and were not reported in earlier works. 
 
This paper is structured as follows: the relativistic and quadratic corrections to one-loop BBR-shift and the thermal two-loop BBR-shift are derived in the section II. The mixing two-loop BBR-shift is brought by the thermal and virtual photon. We would show these results in the section III. The section IV is estimation and discussion.  In the end is the conclusion.

\section{The thermal one-loop and two-loop blackbody-radiation shift of atomic energy-levels}

 \subsection{The finite temperature QED and one-loop blackbody-radiation shift(FTQED)} 
 
 In the laboratory, as well as the universe, the environment contains BBR. The BBR  consists of real photons, which satisfy the Planck distribution function $n_{B}(\omega)=(e^{\beta\omega}-1)^{-1}$ [$\beta=(kT)^{-1}$]. The charged particle in this heated vacuum will interact with these real photons. To study this effect, the finite temperature QED \cite{Donoghue00} was developed. The photon propagator is replaced by the ensemble-averaged photon propagator \cite{Donoghue00} in Feynman gauge
\begin{equation}
\begin{aligned}\label{photon propagator in Feynman gauge}
          iD_{\mu\nu}(k)=-i\frac{g_{\mu\nu}}{k^{2}}-2\pi  n_{B}(\omega)g_{\mu\nu}\delta(k^{2}).
\end{aligned}
\end{equation}
The first term in Eq.(\ref{photon propagator in Feynman gauge}) is the virtual photon propagator, and the second term is the contribution of the BBR. In our paper, we consider the contributions of the thermal (real) and virtual photon separately and distinguish them in different Feynman diagrams (Fig.1). The dash-wave line (thermal photon propagator) connecting the Fermion line means the Fermion emits/absorb the thermal photon and then absorb/emits it. Only Feynman diagrams containing the dash-wave line have the contribution to BBR-shift. In this section, we introduce the one-loop and two-loop Feynman diagrams which only have thermal photon propagators. In the next section, we study the Feynman diagrams containing both wave and dash-wave line (mixing contribution). As the thermal photon is real, it is more convenient to study in Coulomb gauge. The photon propagator in Coulomb gauge can be obtained by using gauge transformation $D_{\mu\nu}(k)\Rightarrow D_{\mu\nu}(k)+k_{\mu}f_{\nu}+f_{\mu}k_{\nu}$ \cite{Lindgren00},  
 
\begin{equation}
\begin{aligned}\label{photon propagator in Coulomb gauge}
          iD_{00}(k)=&\frac{i}{\textbf{k}^{2}},\\
           iD_{ij}(k)=&\frac{id_{ij}}{k^{2}}+2\pi  n_{B}(\omega)d_{ij}\delta(k^{2}),
\end{aligned}
\end{equation}
where $d_{ij}(\omega)=(\delta_{ij}-\frac{k_{i}k_{j}}{\omega^{2}})$.
The first term of the RHS in Eq.(\ref{photon propagator in Coulomb gauge}) is the ordinary virtual photon propagator. The thermal photon propagator is transverse (the second term of the RHS in the second line), because the thermal photon is real.

\begin{figure}[htbp]
  \centering
  \includegraphics[width=4.00in,height=2.50in]{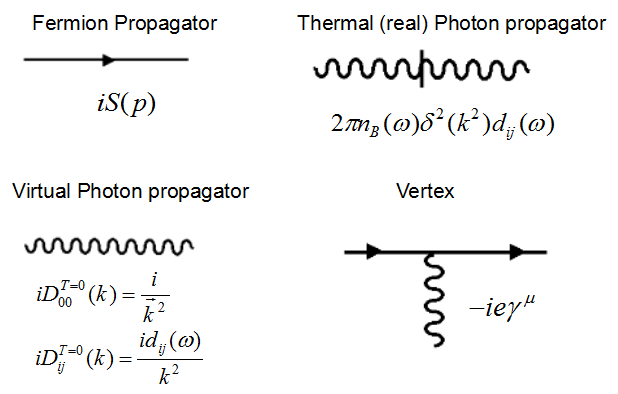}
  \caption{Feynman Rule}
\end{figure}

The thermal one-loop BBR-shift is shown in Fig.2. It leads the thermal mass shift $\delta m(T)=-\frac{\alpha\pi}{3m\beta^{2}}$ \cite{Donoghue00} in the free-field case and also lead the atomic energy-level shift.  In order to calculate it, we replace the internal electron line by the Green function,  
 \begin{equation}\label{Green Function}
 iS(E)=\frac{i}{E-H_{D}(1-i0^{+})},
 \end{equation}
where the $H_{D}$ is Dirac-Hamiltonian. The bra $\langle \psi \mid \sim \psi ^{\dagger}$, so the vertex is  $-ie \alpha^{i}=-ie\gamma^{0}\gamma^{i}$. 
 
\begin{figure}[htbp]
  \centering
  \includegraphics[width=2.00in,height=1.0in]{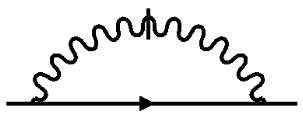}
  \caption{The thermal one-loop correction of BBR-shift}
\end{figure} 
 
The thermal one-loop correction of BBR-shift can be obtained directly in Coulomb gauge by using S-matrix approach \cite{Sucher00,Lindgren00} 
\begin{equation}
\begin{aligned}\label{1loop}
\Delta E_{\textbf{2}} = 
 &e^{2} \int\frac{n_{B}(\omega)d^{3}k}{(2\pi)^{3}2\omega} 
 d_{ij}(\omega)  
 \sum_{a,b}
\langle \psi|\left( \alpha^{i} e^{i\vec{k}\cdot\vec{x}}\right)_{b} 
\left(\frac{1}{E-H_{D}-\omega}+(\omega\rightarrow -\omega)\right)
\left( \alpha^{j} e^{-i\vec{k}\cdot\vec{x}}\right) _{a}|\psi\rangle   ,
 \end{aligned}
\end{equation}
where the operator $(...)_{a,b}$ acts upon the electron $a,b$, and the subscript i in $\Delta E_{i...}$ (here i=2)  means the contribution of the Fig.i in this paper.

\subsection{The nonrelativistic quantum electrodynamics treatment }

The loop corrections to the energy-level in low-Z atom have always been studied by using the nonrelativistic quantum electrodynamics \cite{Jentschura01}. Let's recheck the Eq.(\ref{1loop}). It is the relativistic form. As the philosophy of non-relativistic procedure, the relativistic targets like wave-function $\psi$, vertex $\alpha^{i} exp(ikx)$ and Green function $(E-H_{D}\pm\omega)^{-1}$ could be approximately expressed as a summation of non-relativistic leading terms and relativistic corrections by using Foldy-Wouthuysen (FW) transformation and time-independent perturbation theory sequentially.

According to FW transformation, the transformed Hamiltonian could be expressed as 
\begin{equation}
\begin{aligned}\label{FWTH}
   H_{D}\xrightarrow{FW} H_{FW}&=m+H_{NR}+V_{R}+...  ,\\
   H_{NR}&=\dfrac{\Pi^{2}}{2m}+V-\dfrac{e\sigma\cdot\textbf{B}}{2m},\\
   V_{R}&=-\dfrac{\Pi^{4}}{8m^{3}}
   -\dfrac{e\sigma\cdot(\textbf{E}\times\Pi-\Pi\times\textbf{E})}{8m^{2}}
   +\dfrac{e\nabla\cdot\textbf{E}}{8m^{2}},
\end{aligned}
\end{equation} 
where the $\Pi^{i}=p^{i}-eA^{i}$, $H_{NR}$ is the nonrelativistic 
Hamiltonian, and $V_{R}$ is the leading relativistic correction.
The relativistic energy could also be accordingly written as 
\begin{equation}\label{FWTE}
E=m+E_{NR}+\langle V_{R}\rangle...,
\end{equation} 
where the $E_{NR}$ is the eigenvalue of $H_{NR}$.
In this way, the modified electron propagator is
\begin{equation}
\begin{aligned}\label{FWTG}
   \dfrac{1}{E-H_{D}-\omega} \xrightarrow{FW} & \dfrac{1}{E-H_{FW}-\omega}
   =\dfrac{1}{E_{NR}+\langle V_{R} \rangle-H_{NR}-V_{R}-\omega}\\
  & =\dfrac{1}{E_{NR}-H_{NR}-\omega}+
   \dfrac{1}{E_{NR}-H_{NR}-\omega}(V_{R}-\langle V_{R} \rangle)\dfrac{1}{E_{NR}-H_{NR}-\omega}+....
\end{aligned}
\end{equation}
The second term in the second line is the first-order relativistic correction to the propagator.

 The new wave function is  
\begin{equation}
\begin{aligned}\label{FWTW}
 |\psi\rangle \xrightarrow{FW} &|\varphi_{FW}\rangle=
 |\varphi_{NR}\rangle+|\delta\varphi_{NR}\rangle+..., \\
 &|\delta\varphi_{NR}\rangle= 
 \left( \dfrac{1}{E_{NR}-H_{NR}}\right) ' V_{R}|\varphi_{NR}\rangle,
\end{aligned}
\end{equation}  
where $|\varphi_{NR}\rangle$ is the eigenvector of $H_{NR}$. $|\delta\varphi_{NR}\rangle$ is the leading term of relativistic correction and the prime means the term within the parenthesis is the reduced Green function.

The new Hamiltonian indicates that the electron is coupling with the photon by the current 
\begin{equation}
\begin{aligned}\label{FWTC1}
 \alpha^{i} e^{i\vec{k}\cdot\vec{r}} \xrightarrow{FW} & 
 \dfrac{p^{i}}{m}\left(1+i \textbf{k}\cdot\textbf{r}-\dfrac{(\textbf{k}\cdot\textbf{r})^{2}}{2} \right)
-\dfrac{(i\textbf{k}\times \sigma)^{i}}{2m} 
 \left(1+i \textbf{k}\cdot\textbf{r}\right)+\delta j^{i}... \\
 &=\dfrac{p^{i}}{m}+\sum_{m}iQ^{1im}k^{m}+\sum_{m,n}iQ^{2imn}k^{m}k^{n}+\delta j^{i}...
 ,  
\end{aligned}
\end{equation}
where the first two terms in the first line are the nonrelativistic current operators and their Taylor expansion $e^{i\vec{k}\cdot\vec{r}}=1+i \textbf{k}\cdot\textbf{r}... $. The $\delta j^{i}$ is the relativistic correction to the current operator 
\begin{equation}
\begin{aligned}\label{FWTC2}
 \delta j^{i}=-\dfrac{p^{i}\textbf{p}^{2}}{2m^{3}}
 -\dfrac{Z\alpha (\textbf{r}\times \sigma)^{i}}{2m^{2}r^{3}}
 +\dfrac{i\omega (\textbf{p}\times \sigma)^{i}}{4m^{2}}.  
\end{aligned}
\end{equation} 
They are induced by $V_{R}$. The coefficient $Q$ in the second line are
\begin{equation}
\begin{aligned}\label{QC}
Q^{1im}=&\dfrac{p^{i}r^{m}-\sum_{k}\varepsilon^{imk}\sigma^{k}}{2m},  \\
Q^{2imn}=&\dfrac{p^{i}r^{m}r^{n}-\sum_{k}\varepsilon^{imk}\sigma^{k}r^{n}}{2m}&.  
\end{aligned}
\end{equation}
The most important part in the Eq.(\ref{FWTC1}) is $p^{i}/m \times 1$. We can always assume $e^{i\vec{k}\cdot\vec{r}}=1$, as the BBR's characteristic wave-length $(kT)^{-1}\simeq 10^{-5}m$ is much larger than the atomic radius in the room temperature.

As this approximation procedure we introduced, the thermal one-loop correction is given into the form
\begin{equation}
\begin{aligned}\label{1loopexpand}
\Delta E_{2} = &\delta m(T)+\Delta E_{\textbf{2E1}}+\Delta E_{\textbf{2R}}+\Delta E_{\textbf{2J}}+\Delta E_{\textbf{2Q}}....
\end{aligned}
\end{equation}
The most important part of $\Delta E_2$ are the first two terms $\delta m(T)$, $\Delta E_{\textbf{2E1}}$ , i.e. the thermal mass shift and electric-dipole contribution (the subscript E1 means electric-dipole). They result from the situation that we only take the leading terms in each approximated part.
\begin{equation}
\begin{aligned}\label{1loopE}
&\delta m(T)+\Delta E_{\textbf{2E1}}=
 e^{2} \int\frac{n_{B}(\omega)d^{3}k}{(2\pi)^{3}2\omega} 
 d_{ij}(\omega)
\\&\times
 \sum_{a,b}
\langle \varphi_{NR}|\left( p^{i}\right)_{b} 
\left(\frac{1}{E_{NR}-H_{NR}-\omega}+(\omega\rightarrow -\omega)\right)
\left( p^{j}\right) _{a}|\varphi_{NR}\rangle.   
\end{aligned}
\end{equation} 
The electric-dipole contribution is
\begin{equation}\label{1loopE1}
 \Delta  E_{\textbf{2E1}}=
\frac{4\alpha}{3\pi} \sum_{n,a,b}  \mathcal{P} \int
\frac{E_{\varphi n} n_{B}(\omega)\omega^{3}d\omega}{(E_{\varphi n})^{2}-\omega^{2}} 
\langle \varphi|(r^{i})_{b}|n\rangle
\langle n|(r^{i})_{a}|\varphi\rangle,
 \end{equation}  
where $E_{\varphi n}=E_{\varphi}-E_{n}$, $\mathcal{P}$ is the principal value of the integral. The principal value of integral is induced to avoid the divergence in the resonance region \cite{Solovyev00}. From now on, we no longer distinguish $H_{NR},E_{NR},\varphi_{NR}$ with $H,E,\varphi$, because the difference appears in the higher-order term, which are not currently interested. By considering the AC Stark effect, in quantum mechanics scheme,  Farley \cite{Farley00} first got this formula, while the relativistic corrections to BBR-shift has not been studied before. The $\Delta E_{\textbf{2R}}$ is induced by the relativistic correction of  electron propagator [Eq.(\ref{FWTG})] and wave function [Eq.(\ref{FWTW})] to the Eq.(\ref{1loop})
\begin{equation}
\begin{aligned}\label{1loopR}
\Delta E_{\textbf{2R}} = &e^{2}
\mathcal{P}\int\frac{n_{B}(\omega)d^{3}k}{(2\pi)^{3}2\omega} 
d_{ij}(\omega) \langle \psi|
 \sum_{a,b}
\Bigg\{
\left( \dfrac{p^{i}}{m} \right) _{a}
\dfrac{1}{E-H-\omega}
(V_{R} -\langle V_{R}\rangle)
\dfrac{1}{E-H-\omega}
 \left(\dfrac{p^{j}}{m} \right) _{b}
\\&+
2V_{R} \left( \dfrac{1}{E-H}\right) '
\left( \dfrac{p^{i}}{m} \right) _{a}
\dfrac{1}{E-H-\omega}
\left( \dfrac{p^{j}}{m} \right) _{b}
+(\omega\rightarrow -\omega)\Bigg\}
|\psi\rangle,
\end{aligned}
\end{equation}
where we have used the electric-dipole approximation to the vertices $\gamma^{i} e^{i\vec{k}\cdot\vec{x}} \rightarrow \dfrac{p^{i}}{m} $.

Another relativistic correction to Fig.2 can be obtained by keeping one vertex to be relativistic current [Eq.(\ref{FWTC2})] and other parts (energy, Hamiltonian, wave function and another vertex) are the leading terms. The relativistic current correction $\Delta E_{\textbf{2J}}$ is given
\begin{equation}
\begin{aligned}\label{1loopJ}
\Delta E_{\textbf{2J}} =&  2 e^{2} \mathcal{P}\int\frac{ n_{B}(\omega) d^{3}k}{(2\pi)^{3}2\omega} d_{ij}(\omega)
 \sum_{a,b}
\langle \psi|
\left( \delta j^{i}\right) _{a}
\dfrac{2(E-H)}{(E-H)^{2}-\omega^{2}}
\left(\dfrac{p^{j}}{m}\right) _{b}
|\psi\rangle,  \\
\end{aligned}
\end{equation}
where the subscript J denotes the current $\delta j$.

Equations (\ref{1loopR}) and (\ref{1loopJ}) are the leading terms of the relativistic corrections to BBR-shift.

The contribution of the multi-pole moment can be obtained by the same method.
For example, by substituting the $Q$ coefficients in Eq.(\ref{QC}) into Eq.(\ref{1loop}), the contribution of  the quadratic term is
\begin{equation}
\begin{aligned}\label{1loopq}
\Delta E_{\mbox{2Q}} =& 4 e^{2}\mathcal{P}\int\frac{ n_{B}(\omega)\omega^{3} d\omega}{(2\pi)^{2}} 
\dfrac{4\delta_{ij}\delta_{kl}-\delta_{ik}\delta_{jl}
-\delta_{il}\delta_{jk}}{15}
 \sum_{a,b}
\langle \psi|\bigg[
(Q^{1ik})_{a}\dfrac{2(E-H)}{(E-H)^{2}-\omega^{2}}(Q^{1jl})_{b}
\\&
-(Q^{2ikl})_{a}
\dfrac{2(E-H)}{(E-H)^{2}-\omega^{2}}
\left(  \dfrac{\textbf{p}^{j}}{m}\right)_{b}\bigg]
|\psi\rangle,  \\
\end{aligned}
\end{equation}
where the  subscript Q means quadratic.

\subsection{The thermal two-loop blackbody radiation shift}
In this subsection, we study the thermal two-loop BBR-shift, which can be obtained by adding a thermal photon propagator to  Fig.2. There are three Feynman diagrams (Fig.3) that have contributions to the thermal two-loop blackbody radiation shift.

\begin{figure}[htbp]
  \centering
 \includegraphics[width=5.00in,height=1.20in]{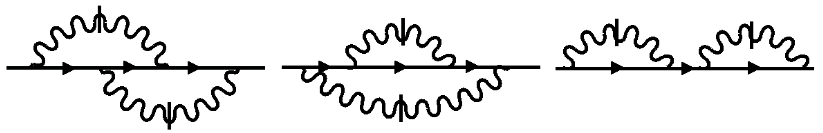}
  \caption{The thermal two-loop contribution of BBR}
\end{figure}

We adopt bound-state S-matrix method to calculate this shift \cite{Sucher00,Lindgren00},
\begin{equation}
\begin{aligned}\label{2loop}
\Delta E_{\textbf{3}} & =
e^{4} \sum_{n_{1}n_{2}n_{3}}\sum_{a,b,c,d}\mathcal{P}\int
\frac{ n_{B}(\omega)d^{3}k}{(2\pi)^{3}2\omega}
\frac{n_{B}(\omega')d^{3}k'}{(2\pi)^{3}2\omega'}\Bigg\{
\\& 
d_{ik}(\omega)d_{jl}(\omega')
\langle \varphi |( \alpha^{l} e^{i\vec{k'}\cdot\vec{x}})_{d} |n_{3}\rangle
\langle n_{3}|(\alpha^{k} e^{i\vec{k}\cdot\vec{x}})_{c} |n_{2}\rangle
\langle n_{2}|( \alpha^{j} e^{-i\vec{k'}\cdot\vec{x}})_{b} |n_{1}\rangle
\langle n_{1}|( \alpha^{i} e^{-i\vec{k}\cdot\vec{x}})_{a} |\varphi \rangle
\\& 
\bigg[\frac{1}{E_{\varphi n_{3}}-\omega'} 
\frac{1}{E_{\varphi n_{2}}-\omega'-\omega} 
\frac{1}{E_{\varphi n_{1}}-\omega}+
\frac{1}{E_{\varphi n_{3}}+\omega'} 
\frac{1}{E_{\varphi n_{2}}+\omega'-\omega} 
\frac{1}{E_{\varphi n_{1}}-\omega}+
\\&
\frac{1}{E_{\varphi n_{3}}-\omega'} 
\frac{1}{E_{\varphi n_{2}}-\omega'+\omega} 
\frac{1}{E_{\varphi n_{1}}+\omega}+
\frac{1}{E_{\varphi n_{3}}+\omega'} 
\frac{1}{E_{\varphi n_{2}}+\omega'+\omega} 
\frac{1}{E_{\varphi n_{1}}+\omega}
\bigg]+
\\&
d_{il}(\omega)d_{jk}(\omega')
\langle \varphi|( \alpha^{l} e^{i\vec{k}\cdot\vec{x}}) _{d}|n_{3}\rangle
\langle n_{3}|(\alpha^{k} e^{i\vec{k'}\cdot\vec{x}}) _{c}|n_{2}\rangle
\langle n_{2}|(\alpha^{j} e^{-i\vec{k'}\cdot\vec{x}}) _{b}|n_{1}\rangle
\langle n_{1}|(\alpha^{i} e^{-i\vec{k}\cdot\vec{x}}) _{a}|\varphi\rangle 
\\& 
\bigg[\frac{1}{E_{\varphi n_{3}}-\omega}
 \frac{2(E_{\varphi n_{2}}-\omega)}{(E_{\varphi n_{2}}-\omega)^{2}-\omega'^{2}} 
 \frac{1}{E_{\varphi n_{1}}-\omega}+
 \frac{1}{E_{\varphi n_{3}}+\omega} 
 \frac{2(E_{\varphi n_{2}}+\omega)}{(E_{\varphi n_{2}}+\omega)^{2}-\omega'^{2}} 
 \frac{1}{E_{\varphi n_{1}}+\omega}\bigg]+
\\&
d_{ij}(\omega)d_{kl}(\omega')
\langle \varphi |(\alpha^{l} e^{i\vec{k'}\cdot\vec{x}})_{d} |n_{3}\rangle
\langle n_{3}|(\alpha^{k} e^{-i\vec{k'}\cdot\vec{x}})_{c} |n_{2}\rangle
\langle n_{2}|( \alpha^{j} e^{i\vec{k}\cdot\vec{x}})_{b} |n_{1}\rangle
\langle n_{1}|( \alpha^{i} e^{-i\vec{k}\cdot\vec{x}})_{a} |\varphi \rangle 
\\& 
\bigg[\frac{2E_{\varphi n_{3}}}{E_{\varphi n_{3}}^{2}-\omega'^{2}} 
\frac{1}{E_{\varphi n_{2}}}
\frac{2E_{\varphi n_{1}}}{E_{\varphi n_{1}}^{2}-\omega^{2}}\bigg]\Bigg\}.
\end{aligned}
\end{equation}

It can be simplified with the electric-dipole approximation by doing the replacement $\alpha^{i} e^{\pm i\vec{k}\cdot\vec{x}}\rightarrow p^{i}/m$. This integral is finite in both ultraviolet regions (suppressed by the Planck 's distribution function $n_{B}(\omega)$) and the infrared region.
 
In this section, we calculate the next-leading-order contribution led by thermal photon of BBR. The condition $E_{ab}\gg kT$ is adequate in low-lying states at room temperature. The BBR-shift Eqs.(\ref{1loopE1})(\ref{1loopR})(\ref{1loopJ})(\ref{1loopq})(\ref{2loop}) can be integrated with respect to $\omega$ by neglecting the $\omega$ in the denominator. Every energy or momentum $\omega,k$ of the thermal photon in the numerator provides a $kT/m\simeq 10^{-8}$ factor after integrating over $\omega$ with $n_{B}(\omega)$. Therefore, by a qualitative estimation, the Eqs.(\ref{1loopE1}).(\ref{1loopq}) and (\ref{2loop}) are proportional to $T^4$, and Eq.(\ref{1loopR}).(\ref{1loopJ}) are proportional to $T^2$. In this paper, the approximation led by $E_{ab}\gg kT$ is always applied to estimate the order of magnitude.

\section{The mixing contributions of the blackbody-radiation shift}
 
\subsection{The contributions of virtual photon to blackbody-radiation shift}
The Feynman diagrams in Fig.4 [(a) The self-energy diagram; (b)The triangle vertex diagram; (c) The vacuum polarization diagram.] are the leading corrections of QED to the atomic energy-level. In last several decades, higher-order terms have been studied \cite{Eides00}. Here we want to study these corrections to the the BBR-shift. The Feynman diagram including the BBR-shift must have the dash-wave line, which is the thermal photon propagator. In this section, we calculate the Feynman diagram containing both the sub-diagram in Fig.4 and a thermal photon propagator. We call these diagrams the mixing contributions of the blackbody-radiation shift. Due to the divergences appearing in the virtual particle loop integral, these mixing contributions are more tangled to be calculated. 

\begin{figure}[htbp]
  \centering
 \includegraphics[width=5.00in,height=1.20in]{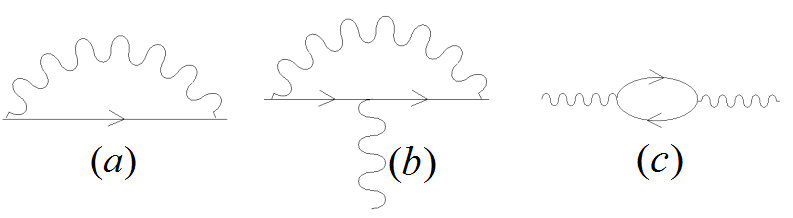}
  \caption{The one-loop Feynman diagram of QED. (a) The self-energy diagram. (b)The triangle vertex diagram. (c)The vacuum polarization diagram. }
\end{figure} 

As the NRQED approach \cite{Jentschura01,Pachucki00}, the contributions of QED effect were classified by the energy of the virtual photon. That is, the high-energy region $\omega\sim m\gg K \sim mZ\alpha$ and the low-energy region $\omega\ll K \sim mZ\alpha$.
These two regions have to be calculated separately. The total result should be independent of the scale $K$. This procedure is proper in the light atoms (Where $m\gg K\sim mZ\alpha\gg m(Z\alpha)^{2}$ and $Z$ is the charge number of the nucleus).

In the following derivation, we neglect the order of $O(K)$ in the high-energy region, as the energy scale is far lager than it.   

The self-energy diagram Fig.4(a) in the free-field \cite{Greiner00} is cancelled with the mass and wave function counterterm, the remainder term is order of $O(K)$, which can be neglected.

The triangle vertex in the free-field \cite{Greiner00}  
\begin{equation}
\begin{aligned}\label{VF}
\Lambda_{\mu}(q)=
-\frac{\alpha}{3\pi}\frac{q^{2}}{m^{2}}
(\textrm{ln}\frac{m}{2K}+\frac{5}{6}-\frac{3}{8})\gamma_{\mu}
+\frac{i}{2m}\frac{\alpha}{2\pi}\sigma_{\mu\nu}q^{\nu}+O(q^{2},K), 
\end{aligned}
\end{equation}
where $q$ is the momentum of the external photon of the triangle diagram (here is the momentum of the Coulomb photon). 
This is the finite part of the triangle vertex of Fig.4(b), and the divergence counterterm has been subtracted. 

The finite part of the vacuum polarization operator\cite{Greiner00} Fig.4(c) is
$\Pi(q)\approx-\alpha q^{2}/(15\pi m^2)$.

The energy-shifts of the hydrogen-like atoms in this section are calculated by using the S-matrix approach first. Then the leading terms are derived by the NRQED. The contributions of self-energy sub-diagram, triangle vertex sub-diagram and vacuum polarization sub-diagram  are obtained sequentially.

\subsection{The contributions of the self-energy sub-diagram to blackbody-radiation shift}

The three Feynman diagrams in the Fig.5 have the self-energy sub-diagram in the Fig.6, which contains infinite diagrams. The thin line in the RHS of Fig.6 is the electron propagator in free-field. As the bound energy and the energy of the thermal photon are both far smaller than the mass of electron, the external electron line of the self-energy sub-diagram is almost on-shell. We can estimate it by the result of QED in free-field. This bound-state self-energy part in Fig.5 could be represented by a series of infinite free-field diagram the Fig.6. In the high-energy region, if the extra electron is on shell, Fig.6(a) is cancelled with the mass and wave function counterterm, the divergence part of Fig.6(b) is cancelled with the vertex counterterm. and each additional Coulomb photon produces an extra factor $Z\alpha$, so the diagrams represented by the ellipsis could also be neglected. The contribution of the self-energy sub-diagram can be replaced by an operator $\Lambda_{\mu}(i\partial) A^{\mu}$ ($\partial$ means replacing the momentum with partial derivative  ). After FW transformation, the operator is given

\begin{equation}
\begin{aligned}\label{SEO}
 \Lambda_{\mu}(i\partial) A^{\mu}=
 V_{SE}+\frac{\alpha}{3\pi m^{2}}
 \textrm{ln}\dfrac{m(Z\alpha)^{2}}{2K} \nabla^2 eA^{0},
\end{aligned}
\end{equation}
where 
\begin{equation}
\begin{aligned}\label{SE1}
 V_{SE}=\frac{\alpha}{3\pi m^{2}}
\left( \textrm{ln}\dfrac{1}{(Z\alpha)^{2}} +\dfrac{5}{6}\right) \nabla^2 eA^{0}+\dfrac{3}{4}\vec{\sigma}\cdot(\nabla eA^{0}\times \vec{p}).
\end{aligned}
\end{equation}
This is the $\textrm{ln}Z\alpha$ term of the self-energy diagram. The remainder term in Eq.(\ref{SEO}) is dependent on the factor $K$, must cancelled with the contribution in the low-energy region. 

\begin{figure}[htbp]
  \centering
 \includegraphics[width=5.00in,height=2.50in]{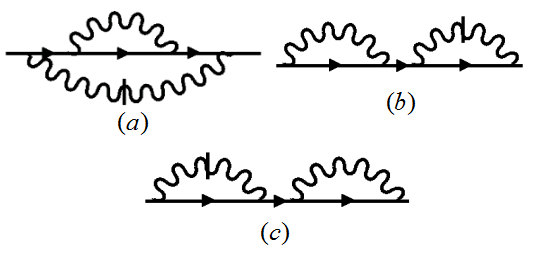}
  \caption{The contributions of the self-energy sub-diagram to BBR shift}
\end{figure}

\begin{figure}[htbp]
  \centering
 \includegraphics[width=5.00in,height=1.20in]{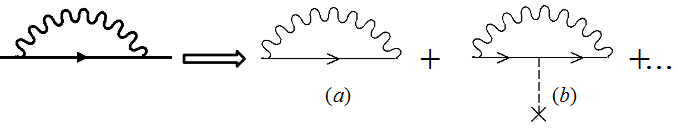}
  \caption{The self-energy sub-diagram contains infinite diagrams, which have an arbitrary number of Coulomb interactions (dashed line). The thin line in the RHS is the electron propagator in the free-field. If the extra electron is on shell, (a) is canceled with the mass counterterm. In the high-energy region, each additional Coulomb photon produces an extra factor $Z\alpha$, so the ellipsis could be neglect. In the low-energy region, all the diagrams in the ellipsis must be included. }
\end{figure}

The leading contribution of Fig.5(a) in the high-energy region can be obtained directly by using Eq.(\ref{SEO}) and the electric-dipole approximation,
\begin{equation}
\begin{aligned}\label{5aH}
\Delta E^{H}_{\textbf{5a}}= &
\dfrac{2\alpha}{3\pi}
\int n_{B}(\omega)\omega d\omega 
\langle \varphi|\Bigg\{
\dfrac{p^{i}}{m}
\dfrac{1}{E-H-\omega}
(V_{SE}+
\frac{\alpha}{3\pi m^{2}}
 \textrm{ln}\dfrac{m(Z\alpha)^{2}}{2K} \nabla^2 eA^{0})
\dfrac{1}{E-H-\omega}
\dfrac{p^{i}}{m}
\\&
+(\omega\rightarrow -\omega) \Bigg\}|\varphi\rangle.
\end{aligned}
\end{equation}
According to the approximation $E_{ab}\gg kT$, this correction is proportional to $T^{2}$.

The contribution of Fig.5(a) in the low-energy region is 

\begin{equation}
\begin{aligned}\label{5aL}
\Delta E^{L}_{\textbf{5a}} & =
\dfrac{4\alpha^{2}}{9\pi^{2}} \int
n_{B}(\omega)\omega d\omega
\omega' d\omega'
\\& 
\langle \varphi |\dfrac{p^i}{m}
\left(
 \dfrac{1}{E-H-\omega}
\dfrac{p^j}{m}\dfrac{1}{E-H-\omega-\omega'}
\dfrac{p^j}{m}\dfrac{1}{E-H-\omega}
+(\omega\rightarrow-\omega)
\right) 
\dfrac{p^i}{m}|\varphi \rangle.
\end{aligned}
\end{equation}
The integral $\omega'$ is linear divergence, as this part divergence is the counterpart of the mass divergence of the self energy sub-diagram. It is can be subtracted by the mass counterterm,
\begin{equation}
\begin{aligned} \label{Substractthemasscounterterm}
  \dfrac{1}{E-H+...-\omega'}\rightarrow \dfrac{1}{E-H+...-\omega'}-\dfrac{1}{-\omega'},
\end{aligned}
\end{equation}   
Where the ellipsis is $\pm\omega$ or $0$.

The total contribution of Fig.5(a) is

\begin{equation}
\begin{aligned}\label{5a}
\Delta E_{\textbf{5a}}= &
\dfrac{2\alpha}{3\pi}
\int n_{B}(\omega)\omega d\omega 
\langle \varphi|\Bigg\{
\dfrac{p^{i}}{m}
\dfrac{1}{E-H-\omega}
V_{SE}
\dfrac{1}{E-H-\omega}
\dfrac{p^{i}}{m}
+(\omega\rightarrow -\omega) \Bigg\}|\varphi\rangle
\\&
+\dfrac{4\alpha^{2}}{9\pi^{2}}
\int n_{B}(\omega)\omega d\omega 
\langle \varphi|\Bigg\{
\dfrac{p^{i}}{m}
\dfrac{1}{E-H-\omega}
\dfrac{p^{j}}{m}
(E-H-\omega)
\textbf{ln}|E-H-\omega|
\\&
\dfrac{p^{j}}{m}
\dfrac{1}{E-H-\omega}
\dfrac{p^{i}}{m}
+(\omega\rightarrow -\omega) \Bigg\}
|\varphi\rangle.
\\&
+\dfrac{2\alpha^{2}}{9\pi^{2}}
\textrm{ln}\dfrac{m(Z\alpha)^{2}}{2K}
\int n_{B}(\omega)\omega d\omega 
\langle \varphi|\Bigg\{
\dfrac{p^{i}}{m}
\dfrac{1}{E-H-\omega}
\dfrac{p^{i}}{m}\dfrac{p^{2}}{m^{2}}
+
\\&
\dfrac{p^{2}}{m^{2}}
\dfrac{p^{i}}{m}
\dfrac{1}{E-H-\omega}
\dfrac{p^{i}}{m}
+(\omega\rightarrow -\omega) \Bigg\}
|\varphi\rangle,
\end{aligned}
\end{equation}
Although the last term in Eq.(\ref{5a}) dependent on the factor $K$, These terms in total contribution in Fig.5 and Fig.8 will be cancelled.    

The leading contribution of Fig.5(b)(c) can be obtained by the similar way,
\begin{equation}
\begin{aligned}\label{5bc}
\Delta E_{\textbf{5bc}}= &
\dfrac{2\alpha}{3\pi}
\int n_{B}(\omega)\omega d\omega 
\langle \varphi|\Bigg\{
V_{SE}\dfrac{1}{E-H}
 \dfrac{p^{i}}{m}\dfrac{1}{E-H-\omega}
\dfrac{p^{i}}{m}+
\dfrac{p^{i}}{m}\dfrac{1}{E-H-\omega}
\dfrac{p^{i}}{m}\dfrac{1}{E-H} V_{SE}
\\&
+(\omega\rightarrow -\omega) \Bigg\}|\varphi\rangle
+\dfrac{4\alpha^{2}}{9\pi^{2}}
\int n_{B}(\omega)\omega d\omega 
\langle \varphi|\Bigg\{
\dfrac{p^{j}}{m}
\left( \dfrac{1}{E-H-\omega}+(\omega\rightarrow -\omega)\right)
\dfrac{p^{j}}{m}
\\&
\dfrac{1}{E-H}
\dfrac{p^{i}}{m}(E-H)\textbf{ln}|E-H|
\dfrac{p^{i}}{m}+
\dfrac{p^{i}}{m}(E-H)\textbf{ln}|E-H|
\dfrac{p^{i}}{m}
\dfrac{1}{E-H}
\\&
\dfrac{p^{j}}{m}
\left( \dfrac{1}{E-H-\omega}+(\omega\rightarrow -\omega)\right)
\dfrac{p^{j}}{m}
\Bigg\}|\varphi\rangle.
\\&
+\dfrac{2\alpha^{2}}{9\pi^{2}}
\textrm{ln}\dfrac{m(Z\alpha)^{2}}{2K}
\int n_{B}(\omega)\omega d\omega 
\langle \varphi|\Bigg\{
\dfrac{p^{i}}{m}
\dfrac{1}{E-H-\omega}
\dfrac{p^{i}}{m}\dfrac{p^{2}}{m^{2}}
+
\\&
\dfrac{p^{2}}{m^{2}}
\dfrac{p^{i}}{m}
\dfrac{1}{E-H-\omega}
\dfrac{p^{i}}{m}
+(\omega\rightarrow -\omega) \Bigg\}
|\varphi\rangle.
\end{aligned}
\end{equation}

There is another contributions of the self-energy sub-diagram to blackbody-radiation shift (Fig.7), which contains thermal self-energy sub-diagram. It is finite. And the low-energy region contribution is

\begin{equation}
\begin{aligned}\label{7L}
\Delta E^{L}_{\textbf{7}} & =
-\dfrac{4\alpha^{2}}{9\pi^{2}} \int
n_{B}(\omega)\omega d\omega
\sum_{a,b,c}
\langle \varphi |\dfrac{p^i}{m}   
|a \rangle \langle a |\dfrac{p^j}{m}  
|b \rangle \langle b |\dfrac{p^j}{m}  
|c \rangle \langle c |\dfrac{p^i}{m}  
|\varphi \rangle \bigg [
\\& 
\dfrac{
E_{\varphi a}(E_{\varphi b}-\omega-E_{\varphi c})\textbf{ln}|E_{\varphi a}|
+(E_{\varphi b}-\omega)(E_{\varphi c}-E_{\varphi a})\textbf{ln}|E_{\varphi b}-\omega|
+E_{\varphi c}(E_{\varphi a}-E_{\varphi b}+\omega)\textbf{ln}|E_{\varphi c}|}
{(E_{\varphi a}-E_{\varphi b}+\omega)
(E_{\varphi b}-\omega-E_{\varphi c})
(E_{\varphi c}-E_{\varphi a})}
\\&
+(\omega\rightarrow -\omega)\bigg ].
\end{aligned}
\end{equation}
We haven't calculate the contribution of Fig.7 in the high-energy region. However, we suppose it is tiny. As it doesn't dependant the factor $K$ which is the UV cutoff.

\begin{figure}[htbp]
  \centering
 \includegraphics[width=2.00in,height=1.30in]{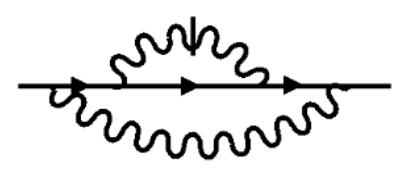}
  \caption{This diagram containing thermal self-energy sub-diagram. It is finite. }
\end{figure}

\subsection{The contributions of the triangle-vertex sub-diagram to blackbody-radiation shift}
The Fig.8 has two Feynman diagrams, which contain a triangle-vertex sub-diagram as Fig.4(b). 

\begin{figure}[htbp]
  \centering
 \includegraphics[width=5.00in,height=1.50in]{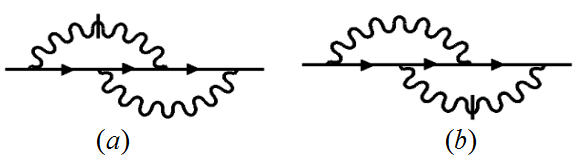}
  \caption{The contributions of the triangle-vertex sub-diagram to BBR shift}
\end{figure}

In the high-energy region the contribution of the triangle-vertex sub-diagram is Eq.(\ref{VF}). As the momentum of the thermal photon (Dash-wave line) is on-shell $q^2=0$, the first term is zero in Eq.(\ref{VF}). Only the second term left. This kind contribution to hydrogen is
\begin{equation}
\begin{aligned}
\Delta E^{H}_{\textbf{8ab}}=&
e^{2}\int \dfrac{n_{B}(\omega)d^{3}k}{(2\pi)^{3}2\omega}d_{ij}
\langle \psi|\bigg[
\alpha^{i}
\dfrac{1}{E-H-\omega}
\dfrac{\alpha}{\pi}\dfrac{i\gamma^{0}\sigma^{i\nu}k_{\nu}}{4m}
+\dfrac{\alpha}{\pi}\dfrac{i\gamma^{0}\sigma^{j\nu}k_{\nu}}{4m}
\dfrac{1}{E-H-\omega}
\alpha^{j}
\\&
+(\omega\rightarrow -\omega)
\bigg]|\psi\rangle,
\end{aligned}
\end{equation}
where $\psi$ is Dirac wave function. As it has the momentum of the thermal photon in the numerator. Comparing with contribution in the low-energy region, which we will derive below, it is suppressed by a $kT/m\simeq 10^{-8}$ factor in the room temperature.
  
  The contributions of the low-energy photon in the Fig.8  are 
\begin{equation}
\begin{aligned}\label{8abL}
\Delta E^{L}_{\textbf{8ab}} & =
\dfrac{4\alpha^{2}}{9\pi^{2}} \int
n_{B}(\omega)\omega d\omega
\omega' d\omega' 
\langle \varphi |\dfrac{p^i}{m}
\bigg[
 \dfrac{1}{E-H-\omega}
\dfrac{p^j}{m}\dfrac{1}{E-H-\omega-\omega'}
\dfrac{p^i}{m}\dfrac{1}{E-H-\omega'}
\\&
+\dfrac{1}{E-H-\omega'}
\dfrac{p^j}{m}\dfrac{1}{E-H-\omega-\omega'}
\dfrac{p^i}{m}\dfrac{1}{E-H-\omega}
+(\omega\rightarrow-\omega)
\bigg] 
\dfrac{p^j}{m}|\varphi \rangle
\\&
=-\dfrac{4\alpha^{2}}{9\pi^{2}} \int
n_{B}(\omega)\omega d\omega
\sum_{a,b,c}
\langle \varphi |\dfrac{p^i}{m}   
|a \rangle \langle a |\dfrac{p^j}{m}  
|b \rangle \langle b |\dfrac{p^i}{m}  
|c \rangle \langle c |\dfrac{p^j}{m}  
|\varphi \rangle \bigg [
\\& 
\dfrac{(E_{\varphi b}-\omega)\textbf{ln}|E_{\varphi b}-\omega|
-E_{\varphi c}\textbf{ln}|E_{\varphi c}|}
{E_{\varphi a}(E_{\varphi b}-\omega-E_{\varphi c})}+
\dfrac{(E_{\varphi b}-\omega)\textbf{ln}|E_{\varphi b}-\omega|
-E_{\varphi a}\textbf{ln}|E_{\varphi a}|}
{E_{\varphi c}(E_{\varphi b}-\omega-E_{\varphi a})}
\\&
+(\omega\rightarrow -\omega)\bigg ]
-\dfrac{4\alpha^{2}}{9\pi^{2}}
\textrm{ln}\dfrac{m(Z\alpha)^{2}}{2K}
\int n_{B}(\omega)\omega d\omega 
\langle \varphi|\Bigg\{
\dfrac{p^{i}}{m}
\dfrac{1}{E-H-\omega}
\dfrac{p^{i}}{m}\dfrac{p^{2}}{m^{2}}
+
\\&
\dfrac{p^{2}}{m^{2}}
\dfrac{p^{i}}{m}
\dfrac{1}{E-H-\omega}
\dfrac{p^{i}}{m}
+(\omega\rightarrow -\omega) \Bigg\}
|\varphi\rangle.
\end{aligned}
\end{equation}  
  It is obvious that all the $\textbf{ln}K$ factor in the Eq.(\ref{5a})(\ref{5bc})(\ref{8abL}) are cancelled exactly.

\subsection{The contributions of the vacuum polarization sub-diagram to blackbody-radiation shift}
The contributions of the vacuum polarization sub-diagram Fig.4(c) to blackbody-radiation shift are shown in the Fig.9. The vacuum polarization operator $\Pi(q)=-\alpha q^{2}/(15\pi m^2)$ \cite{Greiner00}. We can obtain those contributions of Fig.9 by inserting the vacuum polarization operator into the counterpart. In the Fig.9(a)(b)(c), it means inserting $V_{VP}=e\Pi(i\partial)A^{0}$ ($\partial$ means replacing the momentum with partial derivative)  into Fermion line. The result $\Delta E_{\textbf{9(a)(b)(c)}}$ can be obtained by adding $V_{SE}$ with $V_{VP}=e\Pi(i\partial)A^{0}$. The contribution of Fig.9.(d)(e), is supposed to vanish. Because the $q$, which is the 4-momentum of the thermal (real) photon, in the vacuum polarization operator $\Pi(q)=-\alpha q^{2}/(15\pi m^2)$ is on shell.

\begin{figure}[htbp]
  \centering
 \includegraphics[width=5.00in,height=3.00in]{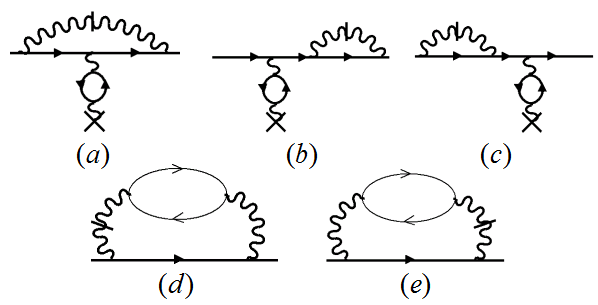}
  \caption{The contributions of the vacuum polarization sub-diagram to BBR shift}
\end{figure}

\subsection{The total contribution of  mixing contributions}
 The total contribution we obtain in Sec.III is
\begin{equation}
\begin{aligned}
\Delta E_{\textbf{5789}}=
\Delta E^{\textbf{ln}Z\alpha}_{\textbf{5789}}+
\Delta E^{\textbf{ln}E}_{\textbf{5789}}+
\Delta E^{H}_{\textbf{8}}.
\end{aligned}
\end{equation}

The most important term is the 
\begin{equation}
\begin{aligned}\label{5789lnza}
&\Delta E^{\textbf{ln}Z\alpha}_{\textbf{5789}}= 
\dfrac{2\alpha}{3\pi}
\int n_{B}(\omega)\omega d\omega 
\langle \varphi|\Bigg\{
\dfrac{p^{i}}{m}
\dfrac{1}{E-H-\omega}
V_{SEVP}
\dfrac{1}{E-H-\omega}
\dfrac{p^{i}}{m}+
\\&
V_{SEVP}\dfrac{1}{E-H}
 \dfrac{p^{i}}{m}\dfrac{1}{E-H-\omega}
\dfrac{p^{i}}{m}+
\dfrac{p^{i}}{m}\dfrac{1}{E-H-\omega}
\dfrac{p^{i}}{m}\dfrac{1}{E-H} V_{SEVP}
+(\omega\rightarrow -\omega) 
\Bigg\}|\varphi\rangle
\end{aligned}
\end{equation}
where
\begin{equation}
\begin{aligned}\label{SEVP1}
 V_{SEVP}=\frac{\alpha}{3\pi m^{2}}
\left( \textrm{ln}\dfrac{1}{(Z\alpha)^{2}} +\dfrac{5}{6}-\dfrac{1}{5}\right) \nabla^2 eA^{0}+\dfrac{3}{4}\vec{\sigma}\cdot(\nabla eA^{0}\times \vec{p}).
\end{aligned}
\end{equation}
It contain the contribution of vacuum polarization. 

The $\Delta E^{\textbf{lnE}}_{\textbf{5789}}$ are the Bethe-logarithm-like terms,
\begin{equation}
\begin{aligned}\label{5789lnE}
&\Delta E^{\textbf{lnE}}_{\textbf{5789}}= 
\dfrac{4\alpha^{2}}{9\pi^{2}}\sum_{a,b,c}
\int n_{B}(\omega)\omega d\omega 
\langle \varphi |\dfrac{p^i}{m}   
|a \rangle \langle a |\dfrac{p^j}{m}  
|b \rangle \langle b |\dfrac{p^k}{m}  
|c \rangle \langle c |\dfrac{p^l}{m}  
|\varphi \rangle
\\&
\Bigg\{
\delta_{il}\delta_{jk}
\Bigg[
\dfrac{(E_{\varphi b}-\omega)\textbf{ln}|E_{\varphi b}-\omega|}
{(E_{\varphi a}-\omega)(E_{\varphi c}-\omega)}
-(E_{\varphi a}-E_{\varphi b}+\omega)^{-1}
(E_{\varphi b}-\omega-E_{\varphi c})^{-1}
(E_{\varphi c}-E_{\varphi a})^{-1}
\\&
[E_{\varphi a}(E_{\varphi b}-\omega-E_{\varphi c})\textbf{ln}|E_{\varphi a}|
+(E_{\varphi b}-\omega)(E_{\varphi c}-E_{\varphi a})\textbf{ln}|E_{\varphi b}-\omega|
\\&
+E_{\varphi c}(E_{\varphi a}-E_{\varphi b}-\omega)\textbf{ln}|E_{\varphi c}|] \Bigg]
+\delta_{ij}\delta_{kl}\Bigg[
\dfrac{E_{\varphi c}\textbf{ln}|E_{\varphi c}|}
{(E_{\varphi a}-\omega)E_{\varphi b}}
+\dfrac{E_{\varphi a}\textbf{ln}|E_{\varphi a}|}
{(E_{\varphi c}-\omega)E_{\varphi b}}\Bigg]
\\&
-\delta_{ik}\delta_{jl}\bigg [ 
\dfrac{(E_{\varphi b}-\omega)\textbf{ln}|E_{\varphi b}-\omega|
-E_{\varphi c}\textbf{ln}|E_{\varphi c}|}
{E_{\varphi a}(E_{\varphi b}-\omega-E_{\varphi c})}+
\dfrac{(E_{\varphi b}-\omega)\textbf{ln}|E_{\varphi b}-\omega|
-E_{\varphi a}\textbf{ln}|E_{\varphi a}|}
{E_{\varphi c}(E_{\varphi b}-\omega-E_{\varphi a})}\bigg ]
\\&
+(\omega\rightarrow -\omega)\Bigg \}.
\end{aligned}
\end{equation} 
We call it BBR-Bethe logarithm.

The less important is 
\begin{equation}
\begin{aligned}
\Delta E^{H}_{\textbf{8}}=&
e^{2}\int \dfrac{n_{B}(\omega)d^{3}k}{(2\pi)^{3}2\omega}d_{ij}
\langle \psi|\bigg[
\alpha^{i}
\dfrac{1}{E-H-\omega}
\dfrac{\alpha}{\pi}\dfrac{i\gamma^{0}\sigma^{i\nu}k_{\nu}}{4m}
+\dfrac{\alpha}{\pi}\dfrac{i\gamma^{0}\sigma^{j\nu}k_{\nu}}{4m}
\dfrac{1}{E-H-\omega}
\alpha^{j}
\\&
+(\omega\rightarrow -\omega)
\bigg]|\psi\rangle,
\end{aligned}
\end{equation}
where $\psi$ is Dirac wave function.

\section{Estimation and Discussion}

Till now, we have studied the one-loop and two-loop BBR-shift. Attribute to the order counting rules of correction terms $\langle p\rangle\sim\langle r\rangle^{-1}\sim mZ\alpha$,$\langle E\rangle \sim m(Z\alpha)^{2} $, we could estimate the order of magnitude of these shifts. The dimension parts of the BBR-shift of low-lying states are listed in Table.I. The approximation $\Delta E \gg kT$ is applied.
 
\begin{table}[htbp]
\caption{The magnitude of BBR-shifts $(Hz)$, which are arranged by the increasing order of $\alpha $ factor. 
$\langle\Delta E_{\textbf{2Ji}}\rangle$ is energy-shift Eq.(\ref{1loopJ}) originating from the $i$th term in Eq.(\ref{FWTC2}).The approximation,  energy-gaps between low-lying states satisfying $\Delta E \gg kT$, is applied.}
\begin{tabular}{cc}
     \hline\hline
     &The magnitude of BBR-shift\\
     \hline
     $\delta m(T)=-\frac{\alpha\pi}{3m\beta^{2}}$& $2.42\times10^{3}\frac{(T)^{2}}{300^{2}}$
     \\
     $\langle\Delta E_{\textbf{2E1}}\rangle
     \sim\frac{1}{Z^{4}m^{3}\alpha^{3}\beta^{4}} $ &
     $\frac{10^{-3}}{Z^{4}}\frac{T^{4}}{300^{4}}$ 
     \\
     $\langle\Delta E_{\textbf{2\textbf{Q}}}\rangle
     \sim\frac{1}{Z^{2}m^{3}\alpha \beta^{4}} $ &
     $\frac{10^{-7}}{Z^{2}}\frac{T^{4}}{300^{4}}$ 
     \\
      $\langle\Delta
     E_{\textbf{3}}\rangle\sim\frac{1}{m^{3}Z^{2}\beta^{4}}$&$\frac{10^{-9}}{Z^{2}}\frac{T^{4}}{300^{4}}$
     \\
     $\langle\Delta E_{\textbf{2J3}}\rangle
     \sim\frac{\alpha}{m^{2}\beta^{3}} $ &
     $10^{-4}\frac{T^{3}}{300^{3}}$ 
     \\
     $\langle\Delta E_{\textbf{2R}}\rangle
     =\langle\Delta E_{\textbf{2J1,2}}\rangle
     \sim\frac{(Z\alpha)^{2}\alpha}{m\beta^{2}}$&$10^{-1}\frac{(ZT)^{2}}{300^{2}}$
     \\    
  $\langle\Delta E_{\textbf{5789}}\rangle \sim\frac{(Z\alpha)^{2}\alpha^{2}}{m\beta^{2}}$&$10^{-3}\frac{(ZT)^{2}}{300^{2}}$\\
     \hline
\end{tabular}
\end{table}

The thermal mass $\delta m(T)$ is the main part of the thermal one-loop correction. However, it is irrelevant to the atomic energy-levels. 

The contribution of the electric dipole to BBR-shift $\Delta E_{\textbf{2E1}}$ is coincide with Farley's work \cite{Farley00}. $\Delta E_{\textbf{2\textbf{Q}}}$ is the quadratic contribution to BBR-shift and proportional to $T^4$. Its action with respect to $\alpha$ and $T$ are the same as the contribution of the magnetic dipole mentioned in Ref\cite{Porsev00}. However, they are not exactly the same, as we only Taylor expand $\exp(ikx)$ rather than multipolar expansion.

The thermal two-loop  BBR-shift $\Delta E_{\textbf{3}}$ is very tiny. It is $T^{2}(mZ\alpha)^{-2}\alpha\simeq 10^{-12}$ weaker than the thermal one-loop correction $\delta m(T)$ in the room temperature. It can be proved that thermal $(i+1)$-loop BBR-shift has an $T^{2i}(mZ\alpha)^{-2i}\alpha^{i}$ factor. Therefore, at current situation, we focus on the diagrams that involve single thermal-photon propagator.

The relativistic corrections to the thermal one-loop BBR-shift are 
$\Delta E_{\textbf{2R,J}}$. The mixing contributions of the BBR-shift, $\Delta E_{\textbf{5,7,9}}$, are induced by the Feynman diagrams have a thermal photon propagator and a virtual one. They are proportional to $(Z\alpha T)^{2}$. As dependence on $Z$ and $T/m$ of these effects are different, the importance of each correction may be changed at various conditions. For example, when $Z=1,T=300K$, the magnitude of $\Delta E_{\textbf{2R}},\Delta E_{\textbf{2J}}$  is larger than $\Delta E_{\textbf{2E1}}$ which is the leading-order term of BBR-shift. As the $T$ (or $1/Z$) decreasing, the former will become more important. Jentschura and his colleagues \cite{Jentschura00} have found a similar correction which is proportional to $(Z\alpha T)^{2}$. The authors added Lamb shift and fine-structure to Eq.(\ref{1loopE1}) by hand, which could be a simple and easy method to apply. However, in this work, the relativistic corrections, such as correction to kinetic energy, spin-orbit interaction (fine structure operator), Darwin term [Eq.(\ref{FWTH})] and the correction to the current [Eq.(\ref{FWTC2})], as well as QED corrections, such as Lamb shift [Eq.(\ref{SEO})], triangle-vertex [Sec.III.C] and vacuum polarization [Sec.III.D] are introduced in the first principle. These effects appear in the ground state of the hydrogen-like atoms, which are absent in the Ref\cite{Jentschura00}.

As discussed above, it is necessary to make an accurate calculation to compare the higher-order corrections (relativistic correction and mixing contribution) with the leading-order term of BBR-shift. The relativistic corrections to BBR-shift ($\Delta E_{\textbf{2R,J}}$) are $\alpha^{-1}\simeq 10^{2}$ larger than the mixing contribution $\Delta E_{\textbf{5,7,9}}$, then we only calculate the relativistic correction to BBR-shift, in the hydrogen atoms. 
. 

Equations (\ref{1loopR}) and (\ref{1loopJ}) are integrated with respect to $\omega$ by neglecting the $\omega$ in the denominator.

\begin{equation}
\begin{aligned}\label{1loopRH}
\Delta E_{\textbf{2R}}\simeq & \dfrac{2 e^{2}}{3}
\mathcal{P}\int\frac{n_{B}(\omega)4\pi\omega^{2}d\omega}{(2\pi)^{3}2\omega}
 \langle \psi|
 \sum_{a,b}
\Bigg\{
\dfrac{ p^{i}}{m}
\dfrac{1}{E-H}
(V_{R} -\langle V_{R}\rangle)
\dfrac{1}{E-H}
\dfrac{ p^{i}}{m}
\\&+
2V_{R} \left( \dfrac{1}{E-H}\right) '
\dfrac{ p^{i}}{m}
\dfrac{1}{E-H}
\dfrac{ p^{i}}{m}\Bigg\}
|\psi\rangle
=-\dfrac{5\pi}{9}\dfrac{\alpha}{m\beta^{2}} 
\langle \psi|\dfrac{\textbf{p}^{2}}{m^{2}}|\psi\rangle
\\
&=-\dfrac{5\pi}{18}\dfrac{\alpha(Z\alpha)^{2}}{mn^{2}\beta^{2}},
\end{aligned}
\end{equation}

\begin{equation}
\begin{aligned}\label{1loopJH}
\Delta E_{\textbf{2J}} \simeq &  \dfrac{4e^{2}}{3}
\mathcal{P}\int\frac{n_{B}(\omega)4\pi\omega^{2}d\omega}{(2\pi)^{3}2\omega}
\langle \psi|
\left( -\dfrac{p^{i}\textbf{p}^{2}}{2m^{3}}
 -\dfrac{Z\alpha (\textbf{r}\times \sigma)^{i}}{2m^{2}r^{3}}
 \right)
\dfrac{1}{E-H}
\dfrac{p^{i}}{m}
|\psi\rangle \\
&=\dfrac{10\pi}{9}\dfrac{\alpha}{m\beta^{2}} 
\left\langle \psi \left| \dfrac{\textbf{p}^{2}}{m^{2}} \right| \psi \right\rangle
=\dfrac{5\pi}{9}\dfrac{\alpha(Z\alpha)^{2}}{mn^{2}\beta^{2}},
\end{aligned}
\end{equation}  
where $n$ is the  principal quantum number. In order to obtain these results, $p^{i}/m=-i[r^{i},H]$ is used, and the third term in Eq.(\ref{FWTC2}) which is suppressed by a factor $kT/m\simeq 10^{-8}$ is neglected (the reason is given at the end of Sec.II). 

The total relativistic correction to BBR-shift is 

\begin{equation}
\begin{aligned}\label{1loopRJH}
\Delta E_{\textbf{2J}}+ \Delta E_{\textbf{2R}}\simeq 
\dfrac{5\pi}{9}\dfrac{\alpha}{m\beta^{2}} 
\left\langle \psi\left|\dfrac{\textbf{p}^{2}}{m^{2}}\right|\psi\right\rangle
=\dfrac{5\pi}{18}\dfrac{\alpha(Z\alpha)^{2}}{mn^{2}\beta^{2}}\simeq \dfrac{0.107}{n^{2}}(\dfrac{ZT}{300})^{2} Hz & 
\end{aligned}
\end{equation}  

  The Table.II shows the electric dipole and relativistic correction to blackbody shift for the ground state of hydrogen and ionized helium, at temperature $T=4K,77K$, and $300K$. The third column is the  electric dipole contribution in Ref\cite{Jentschura00}. The last column is the relativistic correction.
\begin{table}
\caption{The electric dipole and relativistic correction to blackbody shift of the ground state of hydrogen and ionized helium, at temperature $T=4K,77K$,and $300K$. The third column is the  electric dipole contribution in Ref\cite{Jentschura00}. The last column is the relativistic correction
}
\begin{tabular}{cccc}
     \hline\hline
     Nuclear charge & Temperature & $\hslash^{-1}\Delta E_{\textbf{2E1}}$    &$\hslash^{-1}\Delta E_{\textbf{2R+2J}}$  \\
    number  &($\mathbf{K}$) & ($Hz$)\cite{Jentschura00} & ($Hz$) \\
     \hline
   Z=1   & $ 4$  &$-1.22\times 10^{-9}$&$1.90\times 10^{-5}$\\
   Z=1   & $ 77$  &$-1.68\times 10^{-4}$&$7.06\times 10^{-3}$\\
   Z=1   & $ 300$  &$-3.88\times 10^{-2}$&$1.07\times 10^{-1}$\\
   Z=2   & $ 4$  &$-7.65\times 10^{-12}$&$7.63\times 10^{-5}$\\
   Z=2   & $ 77$  &$-1.05\times 10^{-5}$&$2.83\times 10^{-2}$\\
   Z=2   & $ 300$  &$-2.42\times 10^{-3}$&$4.29\times 10^{-1}$\\
 \hline
\end{tabular}
\end{table}
These numerical results indicate that the latter could be more important than the former. That is, at certain situations, the crossing phenomenon between the leading-order correction and relativistic correction to BBR-shift occurs. This phenomenon is probably restricted to the ground state. As the principal quantum number increasing, relativistic correction decreases, since the relativistic effect is related to the square of electron-velocity. While, the electric dipole contribution, which depends on the mean value of $r$, is increasing. As the contribution is close to $1Hz$, which is the magnitude of uncertainties of the clock transition \cite{Porsev00}. We would like to suggest that it is necessary to calculate the relativistic correction to BBR-shift in the multi-electron atoms.
  
\section{Conclusion}
  The one-loop and the two-loop  BBR-shift, are obtained in this work by using S-matrix and NRQED approach. The former consists of the electric dipole contribution, the relativistic corrections, quadratic contribution and so on. The contributions of electric dipole and multi-pole moment are important to the high-lying state, as the mean value of $r$ is large. The relativistic correction is important to the low-lying state, as the mean value of the $p/m$ is large. We calculate relativistic corrections to BBR-shift in the ground state of hydrogen and ionized helium. It is larger than the electric dipole contribution, which is regarded as the leading-order of BBR-shift, and comparable to the magnitude of updated clock transition uncertainties. These effects in the multi-electron atom are suggested to be calculated in the future, which might reduce the uncertainty of the clock transition.
   The two-loop BBR-shift are categorized as the thermal two-loop BBR-shift (two real photon propagator) and the mixing two-loop BBR-shift (one virtual photon propagator and one real photon loop). The thermal two-loop BBR-shift is too tiny to be detected nowadays, though it is finite. The mixing two-loop BBR-shift is also studied, which is $\alpha$ weaker than the relativistic corrections.
   These higher-order corrections were not obtained before. The most important term is the relativistic corrections and mixing two-loop corrections to BBR-shift, which are proportional to $Z^{2}T^2$. As their magnitudes are comparing with the electric dipole contribution to the BBR shift (the leading term, proportional to $Z^{-4}T^4$), these effects may become more important in the low temperature and high-Z hydrogen-like atom.

\newpage

\begin{acknowledgments}
This work was supported by the National Natural Science Foundation of China (No.11674253) and (No.11274246).
\end{acknowledgments}

\newpage

\end{document}